# Superconductivity in the ternary iridium-arsenide BaIr$_2$As$_2$

Xiao-Chuan Wang[1], Bin-Bin Ruan[1], Jia Yu[1], Bo-Jin Pan[1], Qing-Ge Mu[1], Tong Liu[1], Gen-Fu Chen[1,2,3], Zhi-An Ren[1,2,3]*

[1] Institute of Physics and Beijing National Laboratory for Condensed Matter Physics, Chinese Academy of Sciences, Beijing 100190, China

[2] Collaborative Innovation Center of Quantum Matter, Beijing 100190, China

[3] School of Physical Sciences, University of Chinese Academy of Sciences

* Email: renzhian@iphy.ac.cn

**Abstract**

Here we report the synthesis and discovery of superconductivity in a novel ternary iridium-arsenide compound BaIr$_2$As$_2$. The polycrystalline BaIr$_2$As$_2$ sample was synthesized by a high temperature and high pressure method. Crystal structural analysis indicates that BaIr$_2$As$_2$ crystallizes in the ThCr$_2$Si$_2$-type layered tetragonal structure with space group *I4/mmm* (No. 139), and the lattice parameters were refined to be *a* = 4.052(9) Å and *c* = 12.787(8) Å. By the electrical resistivity and magnetic susceptibility measurements we found type-II superconductivity in the new BaIr$_2$As$_2$ compound with a $T_c$ (critical temperature) of 2.45 K, and an upper critical field $\mu_0 H_{c2}(0)$ about 0.2 T. Low temperature specific heat measurements gave a Debye temperature about 202 K and a distinct specific jump with $\Delta C_e/\gamma T_c$ = 1.36, which is close to the value of BCS weak coupling limit and confirms the bulk superconductivity in this new BaIr$_2$As$_2$ compound.

# 1. Introduction

The ThCr$_2$Si$_2$-type materials represent a large number of layered ternary intermetallic compounds consisting of transition metal elements, which have been intensively investigating for their rich physical phenomena mainly originating from the transition elements [1, 2]. The crystal structure of these materials is formed by Cr$_2$Si$_2$-type layers separated by other metal atoms, with Cr atoms forming a square lattice coordinated by Si atoms. In these compounds the lately discovered iron-pnictide high-$T_c$ superconductors AFe$_2$As$_2$ (A = Ca, Sr, Ba, Eu, *etc.*) have the highest superconducting $T_c$ up to 49 K by carrier doping, and superconductivity in them emerges after the suppression of the spin density wave (SDW) magnetic order on the Fe square lattice by means of chemical doping, external pressure or internal crystal defects, *etc*. [3-8]. Besides the iron pnictides, superconductivity was also reported in some iron-free ThCr$_2$Si$_2$-type compounds formed with other transition metals, such as SrNi$_2$As$_2$, CeCu$_2$Ge$_2$, LaRu$_2$As$_2$, BaRh$_2$P$_2$, APd$_2$As$_2$ (A = Ca, Sr and Ba), *etc*. [9-14].

Due to the strong spin-orbit coupling effect, the 5d transition metal compounds have been widely studied for exotic superconductivity. For instance, the pressure-induced superconductivity was found in the Mott insulator 1T-TaS$_2$ [15], the anomalously large superconducting upper critical field exists in noncentrosymmetric CePt$_3$Si [16], the possible high-$T_c$ superconducting anomalies was studied on the surface of Na$_x$WO$_3$ [17], and superconductivity was also found in the geometrically frustrated pyrochlore oxides Cd$_2$Re$_2$O$_7$ and AOs$_2$O$_6$ (A = K, Rb, Cs) [18-21]. For the Mott insulator Sr$_2$IrO$_4$, it was recently proposed to be a possible candidate for realizing high-$T_c$ superconductivity by electron-doping very similar to the cuprates [22]. The intermetallic compound Lu$_5$Ir$_4$Si$_{10}$ is superconducting below 3.9 K and also exhibits a strongly coupled charge density wave (CDW) transition at 83 K [23]. For irridium-containing compounds, superconductivity was also reported in some other materials, such as IrSe$_2$, Cu$_{1-x}$Zn$_x$Ir$_2$S$_4$, CeIrSi$_3$, ScIrP, LaIrP and LaIrAs, *etc*. [24-28], and the ternary ThCr$_2$Si$_2$-type BaIr$_2$P$_2$ and SrIr$_2$As$_2$ were found to be superconducting at around 2.1 K and 2.9 K respectively [29].

In this letter, we report the synthesis of a new $ThCr_2Si_2$-type iridium pnictide compound $BaIr_2As_2$ using the high pressure synthesis method. Bulk superconductivity at 2.45 K was revealed in $BaIr_2As_2$ by the characterizations of resistivity, magnetic susceptibility and specific heat properties.

**2. Experimental details**

The polycrystalline $BaIr_2As_2$ samples were prepared by a high temperature and high pressure synthesis method. Firstly, the BaAs and IrAs precursors were prepared from the respective elements in evacuated silica tubes at 700 °C for 20 hours. Secondly, the fine powders of BaAs, IrAs and Ir were mixed together by the molar ratio of 1:1:1, then ground thoroughly and pressed into small pellets. The pellets were placed in a boron nitride capsule and then sintered at 950 °C and a pressure of 2 GPa for 2 hours using a piston-cylinder type high pressure synthesis apparatus. The obtained samples are black in color and very unstable in air, but stable in inert atmosphere for months. All the sample preparation and handling processes were performed in an argon filled glove box.

The samples were characterized by powder X-ray diffraction (XRD) method on a PAN-analytical X-ray diffractometer with Cu-$K_\alpha$ radiation at room temperature for crystal structure determination and phase analysis. The crystal structure was refined by the Rietveld analysis using GSAS program. The temperature dependence of resistivity was measured by the standard four-probe method using a PPMS system (Physical Property Measurement System, Quantum Design). The temperature dependence of DC magnetic susceptibility was measured using an MPMS system (Magnetic Property Measurement System, Quantum Design).

**3. Results and discussion**

The room temperature XRD patterns for a high pressure synthesized $BaIr_2As_2$ polycrystalline sample are shown in Fig. 1. Minor impurity phase of unreacted Ir was observed from the XRD peaks, which were always inevitable in our experiments due to the low activity of Ir elements. The main diffraction peaks were well indexed based on the layered $ThCr_2Si_2$-type tetragonal crystal structure with the space group *I*4/*mmm* (No. 139), in which the $Ir_2As_2$ layers are separated by the Ba atoms, as depicted in the inset of Fig. 1. The crystal structure of the $BaIr_2As_2$ phase was refined by the Rietveld analysis, and the obtained lattice parameters are a = 4.052(9) Å and c = 12.787(8) Å, which are both much larger than those of $BaIr_2P_2$ from previous report [12, 29]. The chemical composition was also confirmed by the energy-dispersive spectroscopy (EDS) method, with the average atomic ratio of Ba:Ir:As close to 1:2:2 within the instrumental error.

In Fig. 2(a) we show the temperature dependence of resistivity for the $BaIr_2As_2$ sample within the temperature range from 1.8 K to 300 K. The resistivity curve shows a metallic behavior at normal state, and followed by a sharp superconducting transition at low temperature. The onset superconducting critical temperature $T_c$(onset) is 2.45 K, and zero resistivity is attained at 2.35 K, with a narrow transition width of 0.10 K. Fig. 2(b) shows the temperature dependence of resistivity under various magnetic fields below 3 K. The $T_c$(onset) shifts towards the low temperature side with increasing applied magnetic field and at last superconductivity is suppressed at 1.8 K under a field of 800 Oe, indicating a small critical field for $BaIr_2As_2$. The upper critical field $\mu_0 H_{c2}(T)$ was determined by the field dependence of resistive $T_c$(onset) as shown in the inset of Fig. 2(a). The $\mu_0 H_{c2}(T)$ data were fitted by the Ginzburg-Landau theory, given that $H_{c2}(T) = H_{c2}(0)(1 - t^2)/(1 + t^2)$, where $t = T/T_c$ is the reduced temperature and $\mu_0 H_{c2}(0)$ is estimated to be 0.20 T, which is close to that of the isostructural $BaIr_2P_2$ superconductor [12].

The DC magnetization for the $BaIr_2As_2$ sample was measured between 1.8 K and 300 K with both zero-field-cooling (ZFC) and field-cooling (FC) methods under a magnetic field of 10 Oe, as shown in Fig. 3(a). The sharp diamagnetic superconducting transitions were observed at 2.45 K in both ZFC and FC modes, corresponding to the same value as determined from the resistive transition. The estimated superconducting shielding volume fraction from ZFC data is about 92% at 1.8 K, indicating the bulk superconductivity of the sample. Above the superconducting transition to room temperature, the magnetic susceptibility shows paramagnetic behavior and no long-range magnetic order is observed. The isothermal magnetization curve at 1.8 K was measured and presented in Fig. 3(b), and the magnetic hysteresis behavior indicates that the $BaIr_2As_2$ compound is a typical type-II superconductor.

To further clarify its superconducting properties, the low temperature specific heat of $BaIr_2As_2$ sample was measured from 1.8 K to 5 K under magnetic fields of $H$ = 0 and 2 T respectively, as plotted in Fig. 4(a). A sharp specific heat jump happening at 2.4 K supports the bulk nature of superconductivity, and this jump was suppressed under the field of 2 T which exceeds the superconducting upper critical field. At the normal state $C_p/T$ was well fitted with the formula $C_p/T = \gamma + \beta T^2 + \delta T^4$, which gave $\gamma$ = 14.6(0) mJ/mol K$^2$, $\beta$ = 1.17(5) mJ/mol K$^4$ and $\delta$ = 10.7(8) $\mu$J/mol K$^6$. The Debye temperature deduced from $\beta$ is 202 K, which is comparable to that of $BaIr_2P_2$ [12]. The electronic contribution of specific heat $C_e$ was obtained by subtracting the lattice contribution from $C_p$, and the curve $C_e/T$ versus $T$ was plotted in Fig. 4(b). The specific heat jump at $T_c$ was evaluated to be $\Delta C_e/\gamma T_c$ = 1.36, which is close to the BCS weak coupling limit value of 1.43.

In conclusion, we successfully synthesized a new ternary $ThCr_2Si_2$-type $BaIr_2As_2$ compound by a high pressure synthesis method and systematically characterized its physical properties. Low temperature resistivity, magnetization and specific heat measurements revealed that the $BaIr_2As_2$ is a type-II bulk superconductor with a $T_c$ of 2.45 K.

**References:**

[1] Steglich F., Aarts J., Bredl C. D., Lieke W., Meschede D., Franz W. and Schafer H., *Phys. Rev. Lett.*, **43** (1979) 1892.

[2] Shelton R. N., Braun H. F. and Musick E., *Solid State Commun.*, **52** (1984) 797.

[3] Rotter M., Tegel M. and Johrendt D., *Phys. Rev. Lett.*, **101** (2008) 107006.

[4] Sasmal K., Lv B., Lorenz B., Guloy A. M., Chen F., Xue Y. Y. and Chu C. W., *Phys. Rev. Lett.*, **101** (2008) 107007.

[5] Torikachvili M. S., Bud'ko S. L., Ni N. and Canfield P. C., *Phys. Rev. Lett.*, **101** (2008) 057006.

[6] Qi Y. P., Gao Z. S., Wang L., Wang D. L., Zhang X. P. and Ma Y. W., *New J. Phys.*, **10** (2008) 123003.

[7] Qi Y. P., Gao Z. S., Wang L., Zhang X. P., Wang D. L., Yao C., Wang C. L., Wang C. D. and Ma Y. W., *EPL*, **96** (2011) 47005.

[8] Lv B., Deng L., Gooch M., Wei F., Sun Y., Meen J. K., Xue Y. Y., Lorenz B. and Chu C. W., *Proc. Natl. Acad. Sci. U.S.A.*, **108** (2011) 15705.

[9] Bauer E. D., Ronning F., Scott B. L. and Thompson J. D., *Phys. Rev. B*, **78** (2008) 172504.

[10] Jaccard D., Behnia K. and Sierro J., *Phys. Lett. A.*, **163** (1992) 475.

[11] Guo Q., Pan B. J., Yu J., Ruan B. B., Chen D. Y., Wang X. C., Mu Q. G., Chen G. F. and Ren Z. A., *Sci. bull.*, **61** (2016) 921.

[12] Berry N., Capan C., Seyfarth G., Bianchi A. D., Ziller J. and Fisk Z., *Phys. Rev. B*, **79** (2009) 180502.

[13] Anand V. K., Kim H., Tanatar M. A., Prozorov R. and Johnston D. C., *Phys. Rev. B*, **87** (2013) 224510.

[14] Guo Q., Yu J., Ruan B. B., Chen D. Y., Wang X. C., Mu Q. G., Pan B. J., Chen G. F. and Ren Z. A., *EPL*, **113** (2016) 17002.

[15] Sipos B., Kusmartseva A. F., Akrap A., Berger H., Forro L. and Tutis E., *Nat. Mater.*, **7** (2008) 960.

[16] Bauer E., Hilscher G., Michor H., Paul C., Scheidt E. W., Gribanov A., Seropegin Y., Noël H., Sigrist M. and Rogl P., *Phys. Rev. Lett.*, **92** (2004) 027003.

[17] Reich S., Leitus G., Tssaba Y., Levi Y., Sharoni A. and Millo O., *J. Supercon.*, **13** (2000) 855.

[18] Hanawa M., Muraoka Y., Tayama T., Sakakibara T., Yamaura J. and Hiroi Z., *Phys. Rev. Lett.*, **87** (2001) 187001.

[19] Yonezawa S., Muraoka Y., Matsushita Y. and Hiroi Z., *J. Phys.: Condens. Matter*, **16** (2004) L9.

[20] Yonezawa S., Muraoka Y., Matsushita Y. and Hiroi Z., *J. Phys. Soc. Jpn.*, **73** (2004) 819.

[21] Yonezawa S., Muraoka Y. and Hiroi Z., *J. Phys. Soc. Jpn.*, **73** (2004) 1655.

[22] Wang F. and Senthil T., *Phys. Rev. Lett.*, **106** (2011) 136402.

[23] Becker B., Patil N. G., Ramakrishnan S., Menovsky A. A., Nieuwenhuys G. J.

and Mydosh J. A., *Phys. Rev. B*, **59** (1999) 7266.

[24] Guo J. G., Qi Y. P., Matsuishi S. and Hosono H., *J. Am. Chem. Soc.*, **134** (2012) 200001.

[25] Suzuki H., Furubayashi T., Cao G. H., Kitazawa. H., Kamimura A., Hirata K. and Matsumoto T., *J. Phys. Soc. Jpn.*, **68** (1999) 2495.

[26] Sugitani I., Okuda Y., Shishido H., Yamada T., Thamizhavel A., Yamamoto E., Matsuda T. D., Haga Y., Takeuchi T., Settai R. and Ōnuki Y., *J. Phys. Soc. Jpn.*, **75** (2006) 043703.

[27] Okamoto Y., Inohara T., Yamakawa Y., Yamakage A. and Takenaka K., *J. Phys. Soc. Jpn.*, **85** (2016) 013704.

[28] Qi Y. P., Guo J. G., Lei H. C., Xiao Z. W., Kamiya T. and Hosono H., *Phys. Rev. B*, **89** (2014) 024517.

[29] Hirai D., Takayama T., Hashizume D., Higashinaka R., Yamamoto A., Hiroko A. K. and Takagi H., *Physica C*, **470** (2010) S296.

**Figure Captions:**

Figure 1: Powder XRD patterns with Rietveld-refinement for the BaIr$_2$As$_2$ sample. The inset shows a unit cell of the crystal structure.

Figure 2: (a) The temperature dependence of resistivity for the BaIr$_2$As$_2$ sample. The inset shows the upper critical field $\mu_0H_{c2}$. (b) Expanded plots of the resistivity transition under variable magnetic fields.

Figure 3: (a) The temperature dependence of magnetic susceptibility for the BaIr$_2$As$_2$ sample. The inset shows the expanded curve at the transition. (b) The isothermal magnetization curve for the BaIr$_2$As$_2$ sample at 1.8 K.

Figure 4: (a) The specific heat $C_p/T$ versus $T^2$ for 1.8 K < $T$ < 5 K under the fields of 0 and 2 T, and the dotted line represents a fit by $C_p/T = \gamma + \beta T^2 + \delta T^4$. (b) The electronic contribution of specific heat $C_e/T$ versus $T$ for 1.8 K < $T$ < 5 K.

**Fig. 1.**

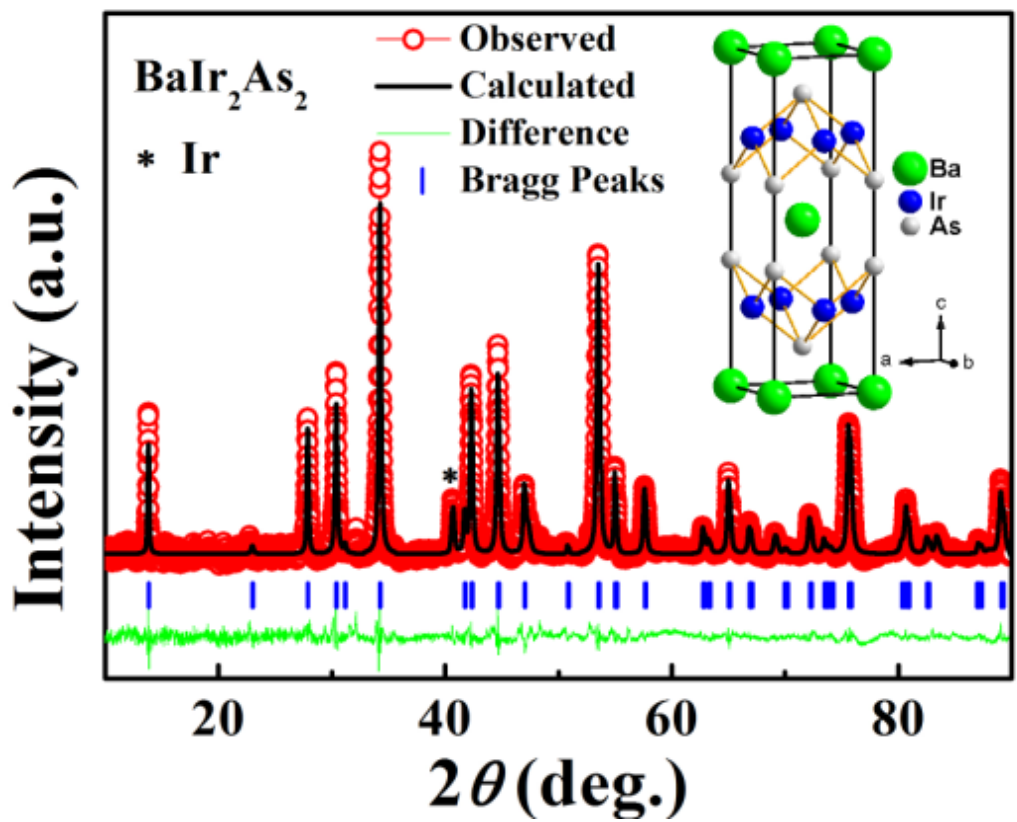

**Fig. 2.**

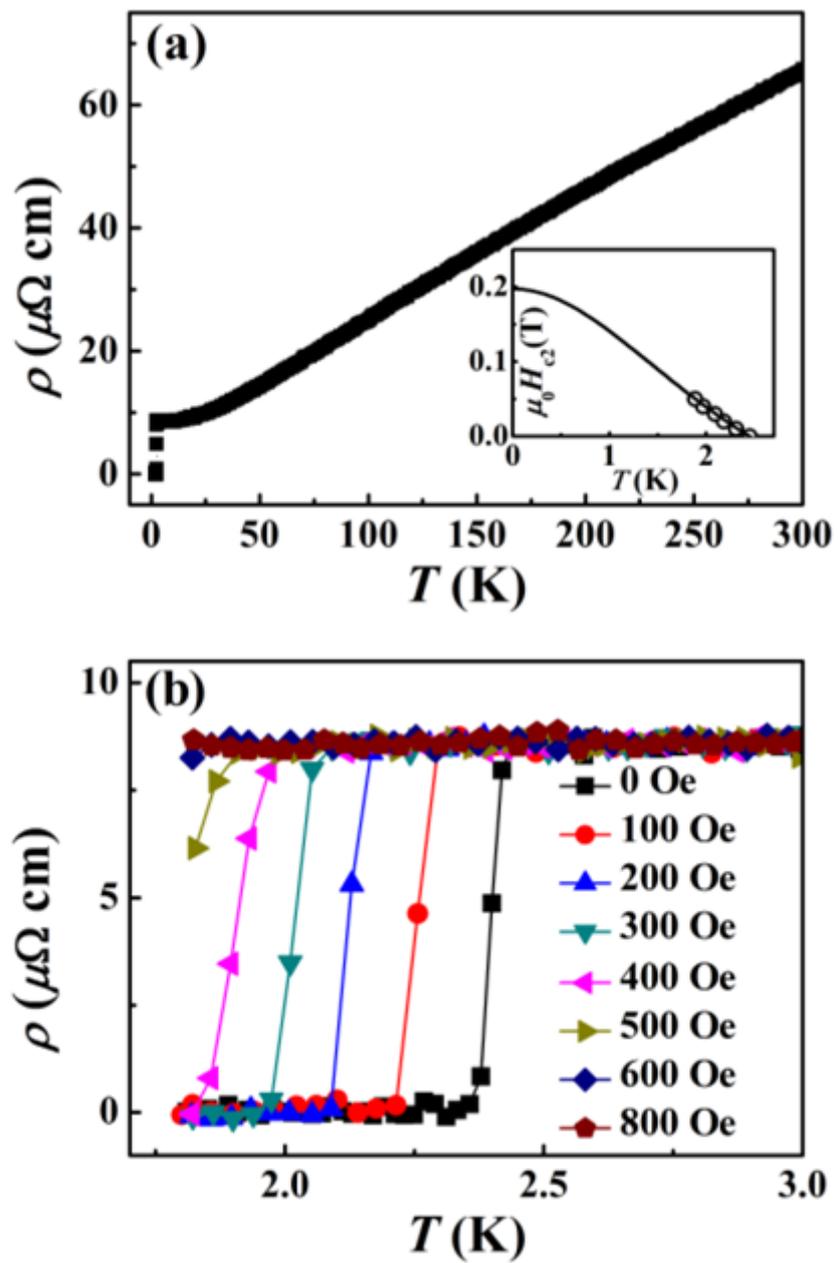

**Fig. 3.**

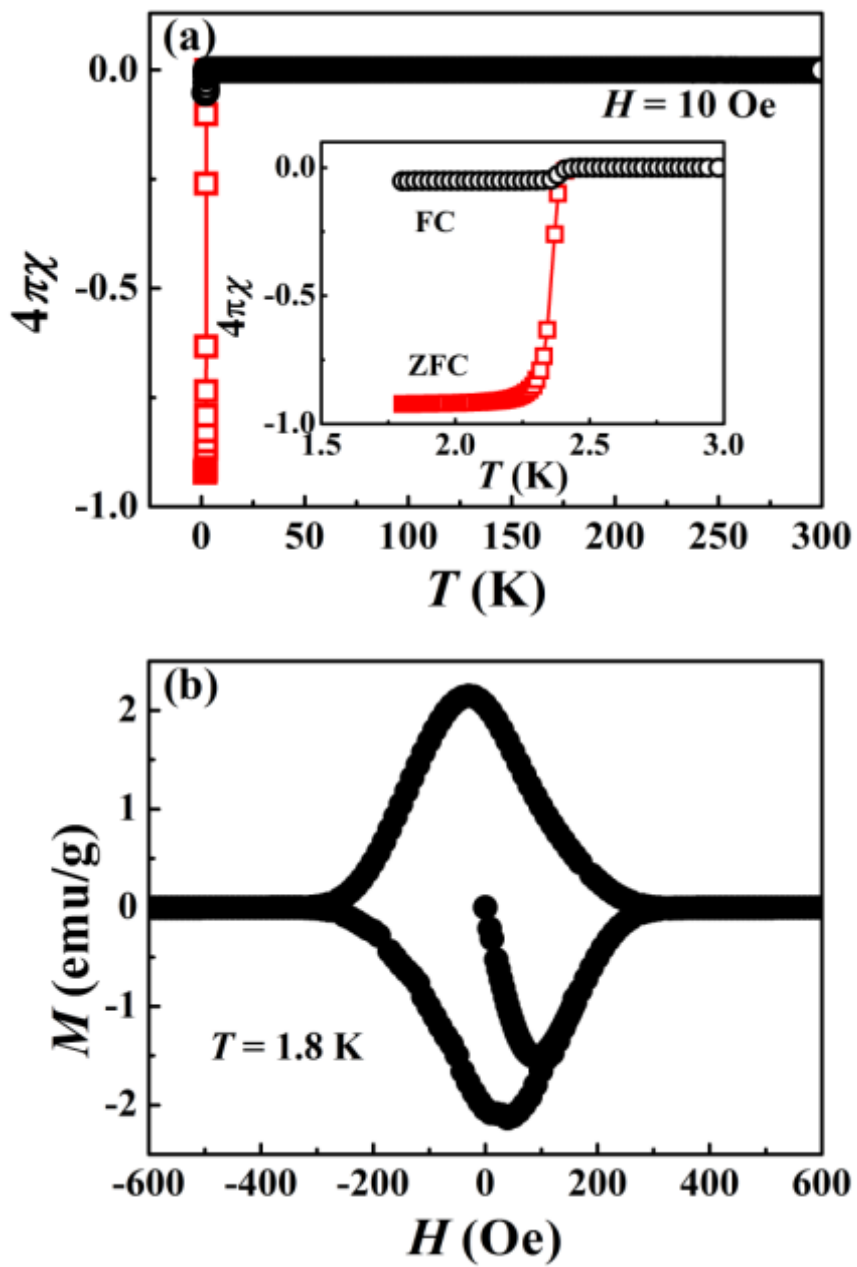

**Fig. 4.**

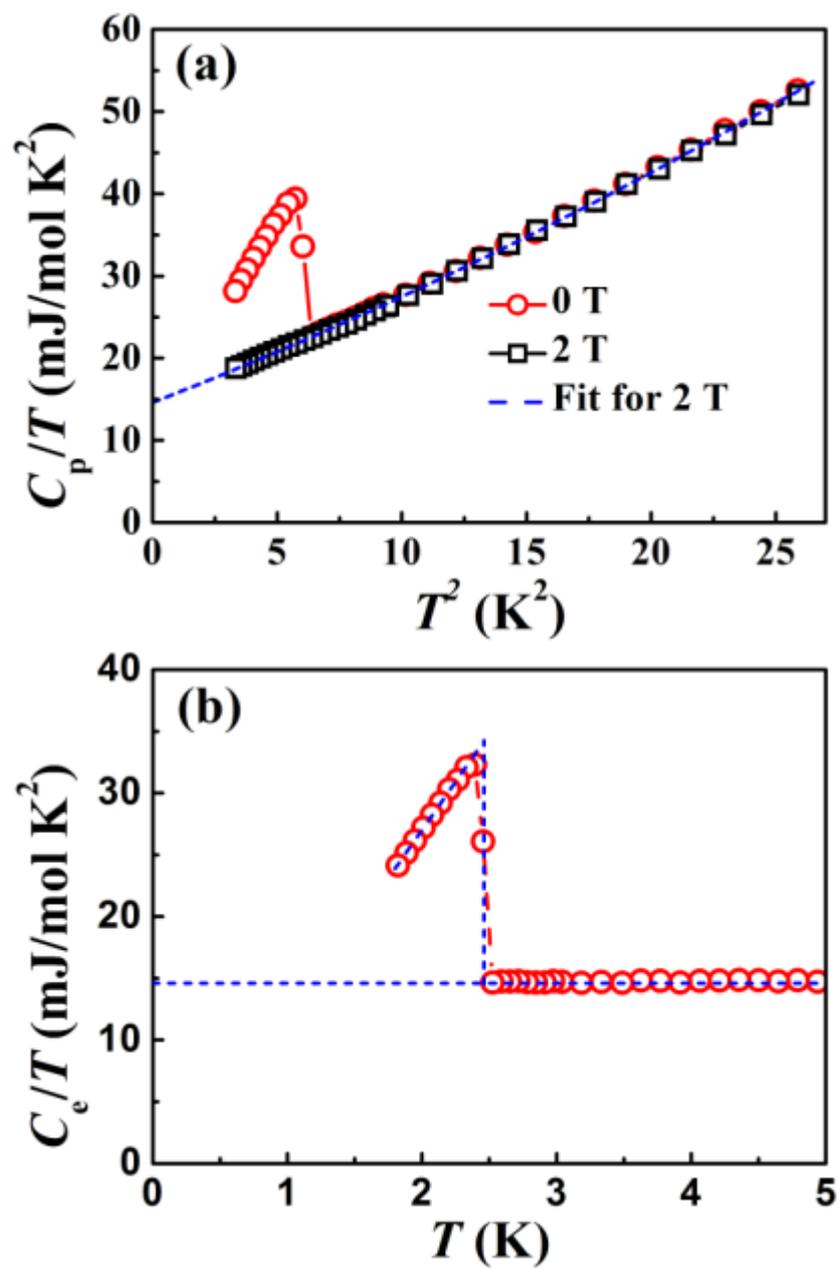